\def\cjaa{Chinese J. Astron. Astrophys.}
\def\apj{ApJ}
\def\aj{AJ}
\def\aap{A\&A}
\def\apjs{ApJS}
\def\mnras{MNRAS}
\def\araa{ARA\&A}

\documentclass[onecolumn,usenatbib]{mn2e}
\usepackage{amssymb}
\usepackage{subfigure}
\usepackage{graphicx,natbib,rotating}

\title[Hard X-ray photon index -- Eddington ratio relation in LLAGNs]
{The anti-correlation between the hard X-ray photon index and the Eddington ratio in LLAGNs}
\author[M. F. Gu and X. Cao]
{Minfeng Gu $^{1}$\thanks{E-mail: gumf@shao.ac.cn}, Xinwu Cao$^{1}$
\\
$^{1}$ Key Laboratory for Research in Galaxies and Cosmology,
Shanghai Astronomical Observatory, Chinese Academy of Sciences, 80\\
Nandan Road, Shanghai 200030, China}

\begin{document}
\pagerange{\pageref{firstpage}--\pageref{lastpage}} \pubyear{...}
\maketitle \label{firstpage}
\begin{abstract}

We find a significant anti-correlation between the hard X-ray photon
index $\Gamma$ and the Eddington ratio $L_{\rm Bol}/L_{\rm Edd}$ for
a sample of Low-Ionization Nuclear Emission-line Regions (LINERs)
and local Seyfert galaxies, compiled from literatures with Chandra
or XMM-Newton observations. This result is in contrast with the
positive correlation found in luminous active galactic nuclei
(AGNs), while it is similar to that of X-ray binaries (XRBs) in
low/hard state. Our result is qualitatively consistent with the
spectra produced from advection dominated accretion flows (ADAFs).
It implies that the X-ray emission of low-luminosity active galactic
nuclei (LLAGNs) may originate from the Comptonization process in
ADAF, and the accretion process in LLAGNs may be similar to that of
XRBs in the low/hard state, which is different from that in luminous
AGNs.

\end{abstract}


\section{Introduction}

Low-luminosity active galactic nuclei \cite[LLAGNs; see][for recent
reviews]{ho08} are believed to be powered by accretion of matter
onto the central supermassive black hole, similarly to powerful
active
galactic nuclei (AGNs). 
The intrinsic faintness, i.e. $L_{\rm Bol}<10^{44} \rm erg~s^{-1}$,
and the low level of activity are the distinctive characteristics of
LLAGNs, which mainly consists of local Seyfert galaxies and
Low-Ionization Nuclear Emission-line Regions (LINERs). In terms of
Eddington luminosity, most of LLAGNs have $L_{\rm Bol}/L_{\rm
Edd}<10^{-2}$, compared to $L_{\rm Bol}/L_{\rm Edd}\sim1$ of
luminous AGNs \citep{pan06}. It still remains unclear whether LLAGNs
are a scaled-down luminosity version of
classical AGNs or objects powered by different physical mechanisms. 
Alternatively, LLAGNs could represent a scaled up version of black
hole binaries in the steady-jet, hard X-ray state, as pointed out
by the scaling relations reported in \citet{fal04}. 
The accretion processes of LLAGNs are still not well understood. For
a sample of 47 local Seyfert galaixes selected from the Palomar
optical spectroscopic survey of nearby galaxies \citep{ho97},
\citet{pan06} found a strong linear correlations between the nuclear
2-10 keV, [OIII]$\lambda5007$, $\rm H_{\alpha}$ luminosities with
the same slope as quasars and luminous Seyfert galaxies, independent
of the level of nuclear activity displayed. The authors thus argued
that the local Seyfert galaxies are consistent with being a
scaled-down version of more luminous AGNs. Most recently, through
revisiting the Spectral Energy Distributions (SED) of 13 nearby
LINERs using recently published, high-angular-resolution data at
radio, ultraviolet (UV), and X-ray wavelengths, \citet{mao07} found
that the SEDs of LINERs are quite similar to those of Seyferts up to
luminosities of $\rm \sim10^{44}~ergs~s^{-1}$, and there is no
evidence for a sharp change in the SED of AGNs at the lowest
luminosities. This result indicates that the thin AGN accretion
disks may persist at low accretion rates, which is also recently
claimed by \cite{pap08} from the fact that the near-IR to X-ray
spectrum of four low-luminosity Seyfert 1 galaxies has the same
shape as that of luminous quasars. However, the radiatively
inefficient accretion flows \citep[RIAFs: an updated version of
ADAF; see, e.g.][references therein]{yua07a} have been invoked to
explain some of the spectral properties observed in many LLAGNs,
such as the lack of the ``big blue bump" \citep{ho99}, and can be
used to explain the SED of several LLAGNs \citep[e.g.][]{chi06}.

X-rays are one of the most direct evidence of nuclear activity and
are, therefore, fundamental to study the accretion processes. The
nature of the X-ray emission has been recently explored for the
samples of LINER galaxies and local Seyfert galaxies from the
observations with Chandra and XMM-Newton telescopes
\citep[e.g.][]{pan04,gon06,pan06}. The high spatial resolution of
Chandra telescope allows the detection of the emission from the
genuine active nuclei, not contaminated by off-nuclear sources
and/or diffuse emisison. Usually, the AGN nature can be defined as
the appearance of nuclei emission in hard X-ray bands. Although the
lack of this feature may likely imply non-AGN contribution, such as
the emission from the diffuse hot gas and/or the starbursts, it does
not necessarily rule out the AGN source nature, since the
non-detection can be due to the heavy absorption. It has been
pointed out that the lack of UV bump in the LINER SED may not be
intrinsic, but simply the effect of obscuration
\cite[e.g.][]{lew03,dud05,mao07}. While the spatial resolution of
XMM-Newton telescope is not enough to separate the off-nuclear
sources and/or diffuse emission, its superior sensitivity and high
throughput provide higher signal-to-noise spectra and the presence
of diffuse thermal emission can be indirectly inferred from the
spectral analysis \cite[e.g.][]{cap06,pan04}. Indeed, the Chandra
and XMM-Newtwon images and spectra of the local Seyfert galaxies
produced in the 0.3-10 keV band have showed a high detection rate
($\sim95$\%) of active nuclei, charcterized, in $\sim60$\% of the
objects, also by the presence of nearby off-nuclear sources and/or,
in $\sim35$\% of the objects, diffuse emission \citep{pan04}. From
an homogeneous analysis of the Chandra X-ray data available for a
sample of 51 LINER galaxies, \citet{gon06} claimed that a high
percentage of LINER galaxies, at least $\approx60\%$, could host AGN
nuclei, and $\sim40\%$ sources lacking a hard nuclear source were
regarded as starburst candidates, because they either lack an
energetically significant AGN or contain highly obscured AGN.
Consistently, both from the Chandra observations, \cite{flo06}
concluded that in about 60\% of 19 LINER sources in their sample an
AGN is present, and \cite{sat05} found that at least 50\% of 82
LINERs harbor a central hard X-ray source consistent with an AGN.

The hard X-ray nuclei emission can well trace the AGN nature in
LLAGNs, however, the origin of the X-ray emission of LLAGNs is still
largely debated. The X-ray emission can be either from the jet
\citep{fal04} or from the RIAF \citep{mer03}. It could even be from
the accretion-flow/hot-corona system of radiative efficient
accretion, as suggested by \cite{pan06} and \cite{mao07} that the
thin AGN accretion disks may still persist at low accretion rates.
From an experimental investigation on the correlation between the
hard X-ray photon index $\Gamma$ with the Eddington ratio $L_{\rm
Bol}/L_{\rm Edd}$ for a sample of luminous AGNs, \cite{she06}
suggested that the hard X-ray photon index can be a good indicator
of accretion rate. In fact, this relation has been explored in
various works, mostly for luminous AGNs
\citep[e.g.][]{lu99,wan04,bia05,gre07} and occasionally for X-ray
binaries (XRBs) \citep{yam05,yua07b}, through which the X-ray origin
and accretion processes were investigated. However, this has never
been explored for LLAGNs. In this paper, we present the relationship
between the hard X-ray photon index $\Gamma$ and the Eddington ratio
$L_{\rm Bol}/L_{\rm Edd}$ for a sample of LLAGNs, with the aim to
constrain the origin of X-ray emission and the accretion processes
in these objects. This paper is organized as follows: The sample is
described in Section 2, the results are given in Section 3 and
Section 4 presents discussions.

\section{The sample}

We started from the samples of LINERs and local Seyfert galaxies of
the Palomar sample \citep{ho97,ho08} and the multiwavelength
catalogue of LINERs (MCL) complied by \cite{car99}. From the Palomar
optical spectroscopic survey of nearby galaxies \citep{ho95},
high-quality spectra of 486 bright ($B_{T}\leq12.5$ mag), northern
($\delta>0^{\circ}$) galaxies have been taken and a comprehensive,
homogeneous catalog of spectral classifications of all galaxies have
been obtained \citep{ho97}. The Palomar survey is complete to
$B_{T}=12.0$ mag and $80\%$ complete to $B_{T}=12.5$ mag. Based on
the relative strength of the optical forbidden lines ([O III]
$\lambda5007$, [O I] $\lambda6300$, [N II] $\lambda6583$ and [S II]
$\lambda\lambda6716,6731$) compared to the hydrogen Balmer lines, H
II nuclei and AGNs were distinguished. Moreover, the ratio [O III]
$\lambda5007/\rm H\beta$ was used to distinguish LINERs ($\rm [O
III]/H\beta<3$) from Seyferts ($\rm [O III]/H\beta\geq3$). From this
survey it has been shown that AGNs are very common ($\sim40\%$) in
nearby galaxies with Seyfert galaxies being $11\%$, pure LINERs
$\sim20\%$ and transition objects (with [O I] strengths intermediate
between those of H II nuclei and LINERs; composite systems having
both an HII region and a LINER component \citep{ho93}) $\sim13\%$ of
all galaxies, respectively. The MCL includes most of the LINERs know
until 1999, providing information on broad band and monochromatic
emission from radio frequencies to X-rays for 476 objects classified
as LINERs. The optical classification was reanalyzed by using the
line-ratio diagrams by \cite{vei87}, and most of the galaxies can be
considered as `pure' LINERs. We searched the literatures for all the
available observations with Chandra or XMM-Newton telescopes. Due to
the fact that the X-ray emission are complex in these objects, and
to ensure that we are exploring the properties of genuine emission
from nuclei AGNs, the source selection were carefully performed.
Thanks to the high spatial resolution of Chandra, the genuine AGNs
are selected from the appearance of the nuclei at hard X-ray band
($>$2 keV). Those sources without the nuclei at hard X-ray band are
excluded in this work, though some of them may have central active
nuclei, which can be heavily obscured \citep{gon06}. To get the hard
X-ray photon index $\Gamma$, the sufficient count rates to perform
the spectral fitting in 2-10 keV band are required. Compton-thick
sources are identified by \cite{pan06} using flux diagnostic
diagrams, which are based on measuring the X-ray luminosity against
an isotropic indicator of the intrinsic brightness of the source to
evaluate the true amount of absorption. Specifically, the flux ratio
of X-ray to [O III]$\rm \lambda5007$ $F_{\rm X}/F_{\rm [O III]}$ and
[O III]$\rm \lambda5007$ to far-infrared $F_{\rm [O III]}/F_{\rm
IR}$ is used to identify the Compton thick sources with $F_{\rm
X}/F_{\rm [O III]}<1$, in addition to identify the AGN-dominant
(non-starburst) emission in sources with $F_{\rm [O III]}/F_{\rm
IR}>10^{-4}$ \citep{pan02,pan06}. In this work, we exclude the
Compton-thick sources identified in \cite{pan06}, since the
Compton-thick absorption may cause the flatter X-ray photon index
and lower absorption when doing spectral fitting in 2-10 keV band,
which may not be real. Moreover, the available measurements of black
hole mass or stellar velocity dispersion are also required.

Finally, we compiled a sample of 55 sources, of which 27 are LINERs
and 28 are local Seyfert galaxies. All the local Seyfert galaxies
are from \cite{ho97} sample, and the LINERs are from either
\cite{ho97} sample or MCL. The photon index from the spectral
fitting and the resulting luminosity in 2-10 keV band of nuclei AGNs
were collected from literatures, of which XMM-Newton data for 20
sources (all are Seyferts) and Chandra data for 33 sources (6
Seyferts and 27 LINERs). In addition, the ASCA data are collected
for 2 Seyfert galaxies (NGC 2639 and NGC 2655) from \cite{pan04} and
\cite{pan06}. Our sample is listed in Table \ref{tbl-1}, in which
column (1) source name, column (2) redshift, column (3) X-ray
satellite used: ``C''=Chandra, ``X''=XMM-Newton, ``A''=ASCA, column
(4) the photon index between 2 and 10 keV, column (5) the 2-10 keV
luminosity in units of $\rm ergs ~s^{-1}$, column (6) the references
of X-ray data, column (7) the mass of central black hole in the
units of solar mass, and column (8) is the references of black hole
masses.

\section{Results}

We collected the black hole masses $M_{\rm BH}$ available from
literatures, which were estimated in different ways from gas,
stellar and maser kinematics to reverberation mapping or inferred
from the mass-velocity dispersion correlations
\citep[e.g.][]{fer00,tre02}. For those $M_{\rm BH}$ estimated with
the mass-velocity dispersion relation different from the
\cite{tre02} relation, we re-calculated $M_{\rm BH}$ with the
\cite{tre02} relation. For those sources without measured $M_{\rm
BH}$ from literatures, we estimate $M_{\rm BH}$ from the stellar
velocity dispersion provided by the Hypercat
database\footnote{Available on-line at
http://www-obs.univ-lyon1.fr/hypercat/} using the \cite{tre02}
relation.
In Table \ref{tbl-1}, we report all these $M_{\rm BH}$ estimates,
the method used to calculate them and the corresponding
references. 
To calculate the bolometric luminosity, we have assumed that $L_{\rm
Bol}/L_{\rm 2-10 keV}\sim30$, which is typical of luminous AGN,
being normally in the range of 25 - 30 \citep{elv94}. It can be seen
from Table \ref{tbl-1} that the X-ray luminosity $L_{\rm 2-10 keV}$
of our sample ranges from $10^{38.6}\rm~erg~s^{-1}$ to
$10^{43.3}~\rm erg~s^{-1}$. Correspondingly, the bolometric
luminosity $L_{\rm Bol}$ varies from $10^{40.1}~\rm erg~s^{-1}$ to
$10^{44.8}~\rm erg~s^{-1}$ with most sources (53 of 55 sources)
lower than $10^{44}~\rm erg~s^{-1}$.

In Fig. \ref{fig1}, we present the relationship between the photon
index $\Gamma$ at 2-10 keV band and the Eddington ratio $L_{\rm
Bol}/L_{\rm Edd}$ for our sample. To compare with luminous AGNs, we
also plot the PG quasars of \cite{she06}. For better comparisons, we
calculated $L_{\rm Bol}=30L_{\rm2-10 keV}$ also for PG quasars,
instead of using optical continuum luminosity in \cite{she06}. It
can be clearly seen that the Eddington ratio $L_{\rm Bol}/L_{\rm
Edd}$ of our sample covers a wide range from $\sim10^{-1}$ going
down to $10^{-6}$, compared to those of most PG quasars $L_{\rm
Bol}/L_{\rm Edd}>10^{-1}$. For most of our sources, the hard X-ray
photon index $\Gamma$ lies in the range of 1.0 - 2.0. We find a
significant anti-correlation between the photon index $\Gamma$ at
2-10 keV band and the Eddington ratio $L_{\rm Bol}/L_{\rm Edd}$ with
a Spearman correlation coefficient of $r=-0.328$ at $\sim99\%$
confidence level for all our LLAGNs. This significant
anti-correlation is found to be slightly improved ($r=-0.517$ at
$\sim99.5\%$ confidence level) when only local Seyfert galaxies are
considered. However, we find no strong correlation ($r=-0.127$ at
$\sim47\%$ confidence level) if we consider LINERs only. This may be
caused by the fact that large scatter exist in photon index of
LINERs, especially those with $\Gamma>2.5$, and LINERs cover
relatively narrower range in $L_{\rm Bol}/L_{\rm Edd}$ than that of
local Seyfert galaxies, as can be clearly seen in Fig. \ref{fig1}.
The significant anti-correlation we found for our LLAGNs is contrast
with the positive correlation for luminous AGNs as clearly shown by
the PG quasars in Fig. \ref{fig1}. 
However, our result is consistent with that of XRBs at low state
\citep{yam05,yua07b}.

For our sample of 55 LLAGNs, a standard $\chi^2$ minimization method
weighted by the errors on $\Gamma$ yields the
following linear relation: 
\begin{equation}
\Gamma=(-0.09\pm0.03)~ \rm log~ (\it L_{\rm Bol}/L_{\rm Edd})+ \rm
(1.55\pm0.07)
\end{equation}
which is shown in Fig. \ref{fig1}. To quantitatively compare with
luminous AGNs, we plot our linear fit to \cite{she06} PG quasars,
i.e. $\Gamma=(0.41\pm0.09)~\rm log~ (\it L_{\rm Bol}/L_{\rm
Edd})+\rm (2.17\pm0.07)$. As an updated version of \cite{she06}
work, we also plot the recent linear fit $\Gamma=(0.31\pm0.01)~\rm
log~ (\it L_{\rm Bol}/L_{\rm Edd})+\rm (2.11\pm0.01)$ presented by
\cite{she08} for a sample of 35 moderate- to high-luminosity radio
quiet AGNs. It should be noted that the \cite{she08} fit is based on
$L_{\rm Bol}$ derived from optical continuum luminosity for a larger
sample. This may cause the differences in two linear fits of
luminous AGNs, which however is consistent with each other within
errors.

\section{Discussions}

In this work, instead of performing a rigorous and homogeneous
spectral analysis, we simply collected the hard X-ray photon index
from the literatures for our sample sources, especially for LINERs
(see Table 1). This may potentially affect the conclusions, since
different authors may use different models to fit the X-ray spectra
(e.g. absorbed power-law, broken power-law, partially absorbed
power-law, power-law plus thermal component, etc.). In principle,
for consistent reasons, the same baseline model should be used.
However, the commonly used method in literatures, e.g., as shown in
\cite{flo06} and \cite{gon06}, is that the different models are
tried to fit the X-ray spectra, and the best fit is usually obtained
with the acceptable $\chi^2$ minimization. For a sample of 51 LINERs
in \cite{gon06}, two models were used to fit Chandra data, i.e. a
single power-law and a single-temperature optically-thin plasma
emission (MEKAL or Raymond Smith (Raymond \& Smith 1977) model). For
each object, five models were attempted: power-law (PL),
Raymond-Smith (RS), MEKAL (ME), PL+RS, and PL+ME.
The best model was chosen as that with the best
$\chi^2$-reduced statistic \cite[see][for details]{gon06}.
As a result, the best fit can be PL for some sources (e.g. NGC
4374), however it can be PL+ME for other sources (e.g. NGC 2681). In
contrast, six models (blackbody, Comptonized blackbody, accretion
disk, hot diffuse gas, power-law and XRB Comptonization) and their
linear combinations were tested in \cite{flo06} in order to find the
most appropriate model for the X-ray spectra of LINERs. Their
spectral fits show that a same model is difficult to well fit all
the LINERs X-ray spectra. On the other hand, the different spectral
fits can be obtained in different works for same Chandra data in
same source. Indeed, seven LINERs in our sample (i.e. NGC 315, NGC
2681, NGC 4374, NGC 4410A, NGC 4457, NGC 4494 and NGC 4552), can be
found with same Chandra data reduced in \cite{gon06} and/or
\cite{flo06}, \cite{sat05}. In these sources, the X-ray spectra can
be well fitted by either single power-law or power-law plus thermal
component, however, the photon index from different works are
consistent with each other within the errors. As a good example, for
NGC 4374, the best fit of \cite{gon06} is power-law model with
$\Gamma=2.07_{-0.20}^{+0.24}$, while \cite{flo06} shows a best fit
of power-law plus MEKAL model with $\Gamma=2.0_{-0.1}^{+0.2}$. For
these seven LINERs, we choose the results with the smallest error in
$\Gamma$. It should be noted that the X-ray data of Seyfert galaxies
are all from \cite{pan04} and \cite{pan06}, in which the homogeneous
analysis was committed. In contrast, LINERs are from various works.
Compared to a significant anti-correlation between $\Gamma$ and
$L_{\rm bol}/L_{\rm Edd}$ in Seyferts, no correlation is found in
LINERs. It is not clear whether the inhomogeneous
spectral analysis smear the correlation in LINERs. 
Nevertheless, it would be important to revisit our results through a
rigorous and homogeneous spectral analysis.
Since our sample data were collected from Chandra and XMM-Newton
telescopes, it is thus interesting to see whether the
anti-correlation still holds using Chandra and XMM-Newton data
separately. We found a significant anti-correlation between $\Gamma$
and $L_{\rm bol}/L_{\rm Edd}$ with $r=-0.553$ at $\sim 99$\%
confidence level for 20 XMM-Newton sources. However, no strong
correlation is found for 33 Chandra sources ($r=-0.177$ at $\sim
68$\% confidence level). Clearly, this result is mainly due to the
fact that 20 XMM-Newton sources are all Seyferts, and 33 Chandra
sources are dominated by LINERs (27 of 33).

The classification of Seyfert/LINER was investigated in many
occasions based on four optical line ratios $\rm [O III]/H\beta$,
$\rm [N II]/H\alpha$, $\rm [S II]/H\alpha$, and $\rm [O I]/H\alpha$
\citep[e.g.][]{hec80,vei87,ho97,kew06}, and the comparisons between
the various classification schemes have also been made, which shown
that the source definition may be different in different
classification scheme \cite[e.g.][]{kew06}. Recently based on a
large number of emission-line galaxies from Sloan Digital Sky
Survey, \cite{kew06} found that Seyferts and LINERs form clearly
separated branches on the optical diagnostic diagrams. They further
derived a new empirical classification scheme, which cleanly
separates Seyferts and LINERs \citep{kew06}. To verify the
classification of our sample, we have tried to re-classify our
sources using their scheme and the line ratios collected from either
\cite{ho97} or \cite{car99}. For our original 27 LINERs and 28
Seyfert galaxies, the re-classification shows that 32 sources are
now LINERs and 23 sources are Seyfert galaxies. Specifically, we
find that all original LINERs are still LINERs except for UGC 08696,
which was regarded as a LINER in \cite{car99} using \cite{vei87}
classification scheme, now is a Seyfert galaxy. This high fraction
of LINERs remaining as LINERs is actually consistent with the
comparisons made by \cite{kew06} between their scheme and others.
Most of LINERs in our sample were defined by \cite{ho97} scheme, and
others by \cite{vei87} scheme in \cite{car99}. It should be noted
that the \cite{ho97} scheme actually is very similar to \cite{vei87}
one. As \cite{kew06} stated that most of galaxies classed as LINERs
in the \cite{ho97} scheme are also LINERs in their scheme. On the
other hand, we find that six original Seyfert galaxies (NGC 2639,
NGC 2655, NGC 3031, NGC 4168, NGC 4579 and NGC 4639) in our sample
may be identified as LINERs with \cite{kew06} scheme. With these new
identifications, we then re-investigate our results. Clearly, the
new identification will not change the anti-correlation for whole
sample of 55 sources. We find that the anti-correlation is still
significant ($r=-0.478$ at $\sim98\%$ confidence level) if we only
consider the 23 Seyfert galaxies newly defined with \cite{kew06}
scheme, and again there is no correlation ($r=-0.127$ at $\sim51\%$
confidence level) for newly classified 32 LINERs. We thus conclude
that the re-classification does not change our results.

\begin{figure}
  \begin{center}
    \includegraphics[width=0.8\textwidth]{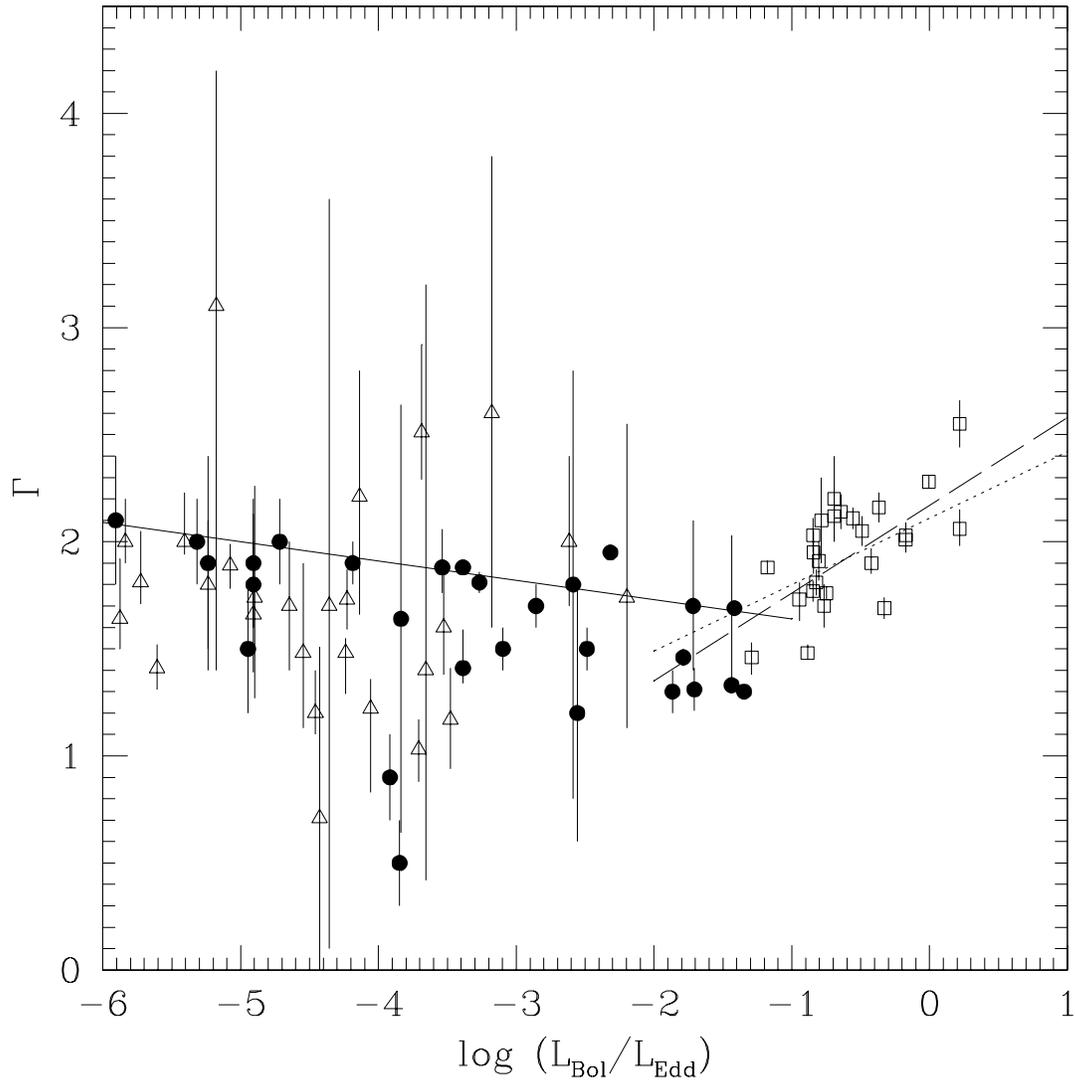}
  \end{center}
\caption{The 2-10 keV photon index versus the Eddington ratio. The
solid circles are for local Seyfert galaxies, while the triangles
are for LINERs. For comparison, PG quasars of Shemmer et al. (2006)
are presented with the rectangles. The solid line is the weighted
linear fit to our LLAGNs sample, while the dashed line is the
weighted linear fit to Shemmer et al. (2006) PG quasars, and the
dotted line is the linear fit to the sample of 35 moderate- to
high-luminosity radio quiet AGNs (Shemmer et al. 2008).}\label{fig1}
\end{figure}

Usually, the bolometric luminosity can be estimated from the
luminosity at a given band by applying a suitable bolometric
correction. The commonly applied bolometric corrections have been
determined from a mean energy distribution calculated from 47
luminous, mostly nearby quasars \citep{elv94}, in which $L_{\rm
Bol}/L_{\rm 2-10 keV}\sim30$. However, as pointed out by
\cite{mar04}, the inclusion of IR emission, which is known to be
re-processed from the ultraviolet (UV), would overestimate the
bolometric luminosity. Moreover, the \cite{elv94} bolometric
corrections were calculated from an X-ray selected sample of
quasars, thus may underestimate the bolometric corrections in the
X-ray band. In addition, the bolometric corrections were derived
from an average quasar template spectrum, thus the dependence on
luminosity of the bolometric corrections cannot be obtained
\citep{mar04}. In calculating the bolometric luminosity $L_{\rm
Bol}$, we have simply assumed $L_{\rm Bol}/L_{\rm 2-10 keV}\sim30$.
Therefore, the bolometric luminosity $L_{\rm Bol}$ in Fig.
\ref{fig1} is just a multiple of the X-ray luminosity. As a matter
of fact, $L_{\rm Bol}$ depends on the shape of the SED, which could
differ among high and low luminosity AGN \citep{ho99,mar04}. Indeed,
the observed $L_{\rm Bol}/L_{\rm 2-10 keV}$ ratio ranges from 3 to
16 \citep{ho99} in LLAGNs, and a lower value of $L_{\rm Bol}/L_{\rm
2-10 keV}\approx10$ was suggested at lower luminosities (typical of
Seyfert galaxies, i.e., $\rm 10^{42}-10^{44} erg~s^{-1}$)
\citep[e.g.][]{fabi04}. The variations of bolometric corrections on
the luminosity was taken into account by \cite{mar04}, in which the
AGN template SED were constructed by using the well-known
anticorrelation between the optical-to-X-ray spectral index and $\rm
2500~\AA$ luminosity. \cite{hop07} also investigated the dependence
of bolometric corrections on the luminosity with similar approach.
Both works shown that the bolometric correction varies among high
and low luminosity AGN. More recently, taking the variations in the
disk emission in the ultraviolet (UV) into account, \cite{vas07}
found evidence for very significant spread in the bolometric
corrections, with no simple dependence on luminosity being evident.
Instead, a more well-defined relationship between the bolometric
correction and Eddington ratio in AGN is suggested, with a
transitional region at an Eddington ratio of $\sim0.1$, below which
the bolometric correction is typically 15-25, and above which it is
typically 40-70. Most recently, the increase in bolometric
correction with Eddington ratio is confirmed and refined by
\cite{vas09} based on simultaneous optical to X-ray SEDs for 29 AGN
from the \cite{pet04} reverberation-mapped sample, with hard X-ray
bolometric corrections of $\kappa_{\rm 2-10 keV} \sim 15-30$ for
$L_{\rm Bol}/L_{\rm Edd}\lesssim 0.1$, $\kappa_{\rm 2-10 keV}
\thickapprox 20-70$ for $0.1\lesssim L_{\rm Bol}/L_{\rm Edd}\lesssim
0.2$ and $\kappa_{\rm 2-10 keV} \sim 70-150$ for $L_{\rm Bol}/L_{\rm
Edd}\gtrsim 0.2$. Since the X-ray luminosity $L_{\rm 2-10keV}$ in
our sample covers over four orders of magnitude, clearly the
dependence of bolometric correction on the luminosity can not be
ignored if it does exist. We then calculated the bolometric
correction using the expression of \cite{hop07} in 2-10 keV band
(see their Equation 2). We find that the bolometric correction
$L_{\rm Bol}/L_{\rm 2-10keV}$ of most sources (51 of 55 sources)
range from 8 to 16, and the average value of whole sample is about
11. Although the new calculated bolometric correction varies from
source to source, we find our results are not changed, i.e. the
anti-correlation between $\Gamma$ and $L_{\rm Bol}/L_{\rm Edd}$ is
still significant. Alternatively, if a $L_{\rm Bol}/L_{\rm 2-10
keV}$ ratio of 10 is considered for all our LLAGN sources, then the
Eddington ratio would be a factor of 3 lower than those in Fig.
\ref{fig1}. Again, this will not change our correlation results. It
should be noted that the more accurate bolometric correction can be
performed when the detailed SED of individual source is constructed.
However, this is not available now.
The adoption of a uniform value of $L_{\rm Bol}/L_{\rm 2-10 keV}$
ratio will introduce dispersion of $\Gamma - L_{\rm Bol}/L_{\rm
Edd}$ relation in Fig. \ref{fig1}, and the influence on the
correlation caused from a uniform bolometric correction needs
further exploration.

While the X-ray emission may be from the Compton process in corona
in luminous AGNs, it is unclear in LLAGNs.
The fundamental plane of supermassive black hole activity derived
from the combination of X-ray, radio luminosity and black hole mass,
were used to constrain the origin of X-ray emission in LLAGNs
\citep{mer03,fal04}. However, the different scenarios were proposed.
\cite{fal04} suggested that the radio through X-ray emission of
LLAGNs is attributed to synchrotron emission from a relativistic
jet, similarly to the scenario proposed for XRBs in their low/hard
state. In contrast, \cite{mer03} suggested a rediatively inefficient
accretion flow as the origin for X-ray emission, and a relativistic
jet for the radio emission. Recently, it is proposed that in some
individual sources the emission from a jet can be responsible for
the X-ray emission \citep[e.g.][]{yua02,fab03,fab04}. While several
pieces of evidence seem to favor an accretion-related X-ray origin
for four LINERs in radio galaxies, a significant (or even dominant)
contribution from the jet in the X-ray regime cannot be excluded for
some objects \citep{gli08}.

The correlation between the X-ray photon index and the Eddington
ratio has been found in various occasions
\citep[e.g.][]{lu99,wan04,she06,gre07}. \cite{she08} updated the
\cite{she06} work by including more highly luminous AGNs, and they
confirmed that the hard X-ray photon index can be a good indicator
of accretion rate. Moreover, the authors claimed that a measurement
of $\Gamma$ and $L_{\rm X}$ can provide an estimate of $L/L_{\rm
Edd}$ and black hole mass with a mean uncertainty of a factor of
$\lesssim3$. There are various explanations for the correlation
found in luminous AGNs. The X-ray emission of luminous Seyfert
galaxies are usually explained by that the emission is produced by a
disk-corona system, where UV soft photon from the accretion disk are
comptonized and up-scattered into the hard X-ray band by a hot
corona above the accretion disk \citep{haa91,haa93}. As the
accretion rate increases, the disk flux to irradiate the corona
increases. This will cause the corona to cool more efficiently
because of Compton cooling, and so the hard X-ray slope will steepen
with the increasing accretion rate \citep{haa91,haa93,wan04}.
Alternatively, \cite{wan04} suggested that the hot corona may become
weak with increasing accretion rate.
\cite{lu99} found two different accretion classes in Seyfert
galaxies and QSOs through investigating the correlation between the
X-ray photon index and the Eddington ratio. The authors argued that
two classes may correspond to ADAF and thin disk accretion,
respectively. While two distinct classes are apparent at the soft
X-ray photon index and Eddington ratio panel, it is not clear for
the case of the hard X-ray photon index due to the lack of data.
Nevertheless, there are strong positive correlations between the
hard X-ray photon index and the Eddington ratio for the objects with
ASCA data, and for only ADAF class objects as well.
To explain the trend of spectrum steepening with increasing
accretion rate in ADAF-class objects, they invoked the two-zone
accretion disk, i.e. an outer standard thin disk and an inner ADAF.
However, we note that all of their ADAF class objects have $\rm
log~(\it L_{\rm Bol}/L_{\rm Edd})> \rm -2.66$, which represent the
luminous ADAF objects compared to our objects.
Most recently from a large sample of 153 radio quiet quasars,
\cite{kel08} confirmed the correlation between $\Gamma_{\rm X}$ and
both $L_{\rm UV}/L_{\rm Edd}$ and $L_{\rm X}/L_{\rm Edd}$, however
they found the dependence is not monotonic. The change in the
disk/corona structure was proposed by the authors as a possible
explanation for the nonmonotonic dependence.

Our study differs from previous work in that we expand the Eddington
ratio down to very low region, $\leq10^{-6}$, which is unexplored
before. The striking result we find is that the anti-correlation
between the hard X-ray photon index and the Eddington ratio of
LLAGNs is apparently contrast with the positive one of luminous
AGNs. This implies that the X-ray emission mechanism of LLAGNs may
be different from that of luminous AGNs. On the other hand, the
anti-correlation is consistent with that of XRBs at low/hard state
\citep{yam05,yua07b}. Through a systematic spectral analysis for 10
XRBs in the low/hard states, using $RXTE$ and $Beppo-SAX$ data,
\cite{yam05} explored the correlation between the hard X-ray photon
index and luminosity. They found a clear anti-correlation both for
the most frequent observed source XTE J1550-564 and for the compiled
data of all 10 XRBs when the 2-200 keV luminosity $L_{2-200}$ is
less than $2\times10^{37}\rm ~erg~s^{-1}$, corresponding to about
1\% of Eddington limit for 10 solar mass black hole. Recently, the
similar anti-correlation was also found at X-ray luminosity below
$\sim 2\%L_{\rm Edd}$ from the combined data of two stellar mass
black hole X-ray sources, XTE J1118+480 and XTE J1550-564
\citep{yua07b}.

Though there are still debates, LLAGNs are usually believed to be
powered by radiatively inefficient accretion flows,
such as ADAFs and their variants 
\citep{nar94}, instead of the standard geometrically thin optically
thick accretion disk typically proposed as the accretion mechanism
acting in the central regions of luminous AGN
\citep{sha73}. 
We find that the qualitative behavior of the correlation of LLAGNs
is consistent with the prediction of the spectra produced from ADAF
model for XRBs in
low state \citep{esi97}. 
At these low accretion rates, the Comptonization of synchrotron
photons by the hot gas in the ADAF constitutes the dominant cooling
mechanism. As the accretion rate increases, the optical depth of the
ADAF goes up causing a corresponding increase in the Compton
$y$-parameter. Consequently, the high-energy part of the spectrum
becomes harder and smoother, and the photon index in the hard X-ray
band hardens \citep[see Fig. 3a of][]{esi97}. In the self-similar
solution of ADAFs \citep{nar95,nar98}, the electron scattering
optical depth $\tau_{\rm es}$ can be expressed as
\begin{equation}
\tau_{\rm es}\simeq24~\alpha^{-1}\dot{m}r^{-1/2}
\end{equation}
where $\alpha$ represents the viscosity parameter \citep{sha73},
$\dot{m}$ is the accretion rate in Eddington units, and $r$ is the
radius in Schwarzschild radii. The inverse Compton spectra of soft
photons by the thermal electrons can be expressed as \citep{ryb79}
\begin{equation}
I_{\nu}\propto\nu^{-\alpha}
\end{equation}
with the spectral index $\alpha=-\rm ln \it \tau_{es}/\rm ln \it A$,
where $I_{\nu}$ is intensity, $A$ is the mean amplification of
photon energy per scattering by the electrons with temperature
$T_{e}$ and is dominantly determined by $T_{e}$. Applying the
$\tau_{es}$ expression in equation (2), we can get the spectral
index $\alpha\propto-\rm log~ \it \dot{m}$, since the electron
temperature $T_{e}$ is only weakly dependent of accretion rate. Our
results are thus qualitatively consistent with the theoretical
expectations of inverse Compton scattering in ADAFs. Similarly,
\citet{yua07b} also predicted an anti-correlation between the X-ray
spectral slope and the Eddington-scaled luminosity based on ADAF
model for objects with low Eddington ratio, in which the X-ray
spectral slope is determined by the Compton $y$-parameter.
In view of the consistence of our result with the ADAF model
predictions, our result seems to suggest that the X-ray emission of
LLAGNs may originate from the Comptonization process in ADAFs.

The electron temperature $T_{e}$ in ADAFs is generally in the range
$10^{9}-10^{10}$ K, and depends only weakly on the accretion rate
\citep{nar98}. Therefore, the X-ray emission of LLAGNs in our
concerned energy band 2-10 keV can be from both the Comptonization
and bremsstrahlung emission. The thermal bremsstrahlung spectrum is
cut off at $\nu_{\rm cutoff}\sim kT_{e}/h$, thus, $\nu_{\rm cutoff}$
is in the range from several tens to hundreds keV for the plasma
with electron temperature $T_{e}\sim10^{9}-10^{10}$ K. This implies
that the spectral index of bremsstrahlung emission from ADAFs is
around $\alpha\sim0~ (\Gamma=\alpha+1\sim1)$ in 2-10 keV, which
however is in contrast with the observations (see Fig. \ref{fig1}).
Therefore, we speculate that Comptonization is dominant over the
bremsstrahlung emission in 2-10 keV for our LLAGNs.

While our sample is optically defined and selected, some sources can
be defined as radio galaxies based on the radio structure, such as
NGC 1275, NGC 315, NGC 4261 and NGC 6251 being regarded as FR I
radio galaxies \citep{fan74}. It thus conceivable that the X-ray
emission can be from jet, which previously has been proposed for
general LLAGNs \citep{fal04}. Based on the analysis of time-averaged
spectra combined with model-independent information from X-ray
temporal and spectral variability, and with inter-band information
for four LINERs hosted by radio galaxies, \cite{gli08} claimed that
LINERs represent a very heterogeneous class, which may range from
intrinsically low-luminosity objects (in terms of Eddington ratio),
to objects potentially strongly absorbed, to brighter objects with
properties more in line with Seyfert galaxies. From an investigation
on the X-ray emission mechanism using the coupled jet-ADAF model for
a sample of eight FR I radio galaxies, \citet{wu07} claimed that the
X-ray emission of FR Is are complex from source to source, and X-ray
can be dominantly from ADAF for luminous sources, and mainly from
jet for the least luminous sources, however, can be a mixture of
ADAF and jet for moderate luminosity sources. More generally for
LLAGNs, \cite{yua05} predicted that the X-ray emission is dominated
by the emission from the ADAF when the X-ray luminosity of a system
is above some critical value, however it will be dominated by the
jet when it is below the critical value. The critical value of
$L_{\rm 2-10 keV}$ is claimed to be $\sim 10^{-6} L_{\rm Edd}$ to
$10^{-7} L_{\rm Edd}$, which varies from source to source. Indeed,
the X-ray emission of FR I galaxies NGC 315 \citep{wu07} and NGC
4261 \citep{gli03} were claimed to be dominated by ADAFs since the
X-ray luminosity is relatively high. On the other hand, the jet
origin X-ray emission was proposed for NGC 4374 \citep{wu07} and NGC
4594 \citep{pel03} due to the low X-ray luminosity $<10^{-7}L_{\rm
Edd}$. We find that five sources (one Seyfert galaxies and four
LINERs) in our sample have $L_{\rm 2-10~ keV}<10^{-7}L_{\rm Edd}$,
therefore, the jet contribution in these sources probably can not be
ignored based on the predictions of \cite{yua05}. However, these
five sources well follow the general trend of $\Gamma - L_{\rm
Bol}/L_{\rm Edd}$ in Fig. \ref{fig1}. This implies that the X-ray
emission in these sources may also be dominated by ADAFs as other
sources in our sample, though the jet emission can not be excluded.
Alternatively, to exclude the possible contamination of jet-origin
X-ray emission in low Eddington ratio LINERs and due to the fact
that LINERs represent a very heterogeneous class
\cite[e.g.][]{gli08}, we retry the correlation analysis for LINERs
after excluding four LINERs with $L_{\rm 2-10~ keV}<10^{-7}L_{\rm
Edd}$, of which a jet contribution in X-ray emission is expected
\citep{yua05,wu07}. However, the correlation between $\Gamma$ and
$L_{\rm bol}/L_{\rm Edd}$ is still lack for remaining 23 LINERs
($r=-0.129$ at $\sim44\%$ confidence level).

The RIAF model is an updated version of the ADAF model, in which the
winds are considered and a relatively high $\delta$ (the fraction of
the energy directly heating the electrons to the total viscously
dissipated energy in the accretion flow) is adopted compared with
the ADAF model. The detailed physics of the putative winds is still
quite unclear for RIAF model, and therefore, our discussions is
mainly based on the well studied ADAF model. Since the general
observational features and structures of the ADAFs and RIAFs are
quite similar, the main conclusions of this work will not be altered
with the RIAF model.

In summary, we propose that the X-ray emission of LLAGNs may
originate from the Comptonization process in ADAFs, and LLAGNs may
be similar to XRBs at the low/hard state, where the accretion rate
is low, and radiation from accretion flow is inefficient. The
detailed fitting of the spectra of individual LLAGNs with ADAF-jet
models should further clarify the mechanism of X-ray emission, which
is beyond the scope of this work.

\section*{Acknowledgments}

We thank the anonymous referee for insightful comments and
constructive suggestions. We thank Feng Yuan, Qingwen Wu and Ramesh
Narayan for the helpful comments and discussions. We acknowledge the
usage of the HyperLeda database (http://leda.univ-lyon1.fr). This
work is supported by National Science Foundation of China (grants
10633010, 10703009, 10833002, 10773020 and 10821302), 973 Program
(No. 2009CB824800), and the CAS (KJCX2-YW-T03).


{}

\clearpage

\begin{table*}\centering\begin{minipage}{140mm}
\caption{The sample. \label{tbl-1}}
\begin{tabular}{lccccccc}
\hline\hline Name & $z$ & Sat & $\Gamma$ (2-10 keV) & log $L_{\rm
2-10~ keV}$ &
refs. & log $M_{\rm BH}^{*}$ & refs. \\
\hline
&&&Seyfert galaxies&&&\\
\hline
NGC 1275  &   0.017559  & X &  1.95$\pm0.01$  &   42.83 &1 &  8.51  &  2 \\
NGC 2639  &   0.011128  & A &  1.64$\pm1.00$  &   40.82 &1 &  8.02  &  2 \\
NGC 2655  &   0.004670  & A &  1.20$\pm0.60$  &   41.85 &1 &  7.77  &  2 \\
NGC 2685  &   0.002945  & X &  0.50$\pm0.20$  &   39.94 &1 &  7.15  &  2 \\
NGC 3031  &  -0.000113$^{a}$  & X &  1.90$\pm0.10$  &   40.25 &1 &  7.80$^{b}$  &  2 \\
NGC 3147  &   0.009407  & C &  1.88$_{-0.12}^{+0.18}$  &   41.89 &1 &  8.79  &  2 \\
NGC 3227  &   0.003859  & X &  1.50$\pm0.10$  &   41.74 &1 &  7.59$^{c}$  &  2 \\
NGC 3486  &   0.002272  & X &  0.90$\pm0.20$  &   38.86 &1 &  6.14  &  2 \\
NGC 3516  &   0.008836  & C &  1.31$\pm0.10$  &   42.29 &1 &  7.36$^{c}$  &  2 \\
NGC 3941  &   0.003095  & X &  2.10$\pm0.30$  &   38.88 &1 &  8.15  &  2 \\
NGC 4051  &   0.002336  & C &  1.33$_{-0.03}^{+0.70}$  &   41.31 &1 &  6.11$^{c}$  &  2 \\
NGC 4138  &   0.002962  & X &  1.50$\pm0.10$  &   41.29 &1 &  7.75  &  2 \\
NGC 4151  &   0.003319  & X &  1.30$\pm0.02$  &   42.47 &1 &  7.18$^{c}$  &  2 \\
NGC 4168  &   0.007388  & X &  2.00$\pm0.20$  &   39.87 &1 &  7.95  &  2 \\
NGC 4258  &   0.001494  & C &  1.41$_{-0.07}^{+0.18}$  &   40.86 &1 &  7.61$^{d}$  &  2 \\
NGC 4388  &   0.008419  & C &  1.70$_{-0.30}^{+0.40}$  &   41.72 &1 &  6.80  &  2 \\
NGC 4395  &   0.001064  & X &  1.30$\pm0.10$  &   39.81 &1 &  5.04$^{b}$  &  2 \\
NGC 4477  &   0.004520  & X &  1.90$\pm0.30$  &   39.65 &1 &  7.92  &  2 \\
NGC 4501  &   0.007609  & X &  1.50$\pm0.30$  &   39.59 &1 &  7.90  &  2 \\
NGC 4565  &   0.004103  & X &  1.80$\pm0.20$  &   39.43 &1 &  7.70  &  2 \\
NGC 4579  &   0.005067  & C &  1.88$\pm0.03$  &   41.03 &1 &  7.78  &  2 \\
NGC 4639  &   0.003395  & X &  1.81$\pm0.05$  &   40.22 &1 &  6.85  &  2 \\
NGC 4698  &   0.003342  & X &  2.00$\pm0.20$  &   39.16 &1 &  7.84  &  2 \\
NGC 4725  &   0.004023  & X &  1.90$\pm0.50$  &   38.89 &1 &  7.49  &  2 \\
NGC 5033  &   0.002919  & X &  1.70$\pm0.10$  &   41.08 &1 &  7.30  &  2 \\
NGC 5273  &   0.003549  & X &  1.46$\pm0.04$  &   41.36 &1 &  6.51  &  2 \\
NGC 5548  &   0.017175  & X &  1.69$\pm0.01$  &   43.25 &1 &  8.03  &  2 \\
NGC 7479  &   0.007942  & X &  1.80$\pm1.00$  &   41.12 &1 &  7.07  &  2 \\
\hline
&&&LINERs&&&\\
\hline
NGC 0315  & 0.016485 & C &  1.48$_{-0.19}^{+0.07}$ &  41.64 &3 &  9.24 &  4  \\
UGC 08696 & 0.037780 & C &  1.74$_{-0.61}^{+0.81}$ &  42.18 &3 & 7.74 &  5  \\
IC 1459   & 0.006011 & C &  1.89$_{-0.11}^{+0.10}$ &  40.56 &3 & 9.00$^{b}$ &  6  \\
NGC 1553  & 0.003602 & C &  1.20$_{-0.10}^{+0.20}$ &  40.18 &7 & 8.00 &  7  \\
3C 218    & 0.054878 & C &  1.17$_{-0.23}^{+0.24}$ &  42.15 &4 & 8.99 &  4  \\
NGC 266   & 0.015547 & C &  1.40$_{-0.98}^{+1.80}$ &  40.88 &8 & 7.90$^{e}$ &  9  \\
NGC 2681  & 0.002308 & C &  1.74$_{-0.47}^{+0.52}$ &  38.94 &3 & 7.20 &  7  \\
NGC 3125  & 0.003712 & C &  2.00$_{-0.30}^{+0.40}$ &  39.67 &10 & 5.65 &  10  \\
NGC 3169  & 0.004130 & C &  2.60$_{-1.00}^{+1.20}$ &  41.41 &8 & 7.95 &  6  \\
NGC 3226  & 0.003839 & C &  2.21$_{-0.55}^{+0.59}$ &  40.74 &8 & 8.24 &  6  \\
NGC 3718  & 0.003312 & C &  1.48$_{-0.35}^{+0.42}$ &  40.06 &4 & 7.97 &  4  \\
NGC 4143  & 0.003196 & C &  1.66$_{-0.27}^{+0.47}$ &  40.04 &8 & 8.31 &  6  \\
NGC 4261  & 0.007465 & C &  0.71$_{-0.71}^{+0.80}$ &  41.15 &4 & 8.94 &  4  \\
NGC 4278  & 0.002165 & C &  1.64$_{-0.14}^{+0.28}$ &  39.96 &8 & 9.20$^{b}$ &  6  \\
NGC 4374  & 0.003536 & C &  2.00$_{-0.10}^{+0.20}$ &  39.60 &7 & 8.80 &  7  \\
NGC 4410A & 0.024190 & C &  1.73$\pm0.14$ &  41.24 &4 & 8.83 &  4  \\
NGC 4457  & 0.002942 & C &  1.70$\pm0.30$ &  38.99 &7 & 7.00 &  7  \\
NGC 4494  & 0.004483 & C &  1.80$\pm0.30$ &  39.00 &7 & 7.60 &  7  \\
NGC 4548  & 0.001621 & C &  1.70$_{-1.60}^{+1.90}$ &  39.79 &8 & 7.51 &  6  \\
NGC 4552  & 0.001071 & C &  1.81$_{-0.10}^{+0.24}$ &  39.41 &3 & 8.50 &  7  \\
NGC 4594  & 0.003639 & C &  1.41$_{-0.10}^{+0.11}$ &  40.07 &3 & 9.04$^{b}$ &  6  \\
NGC 4736  & 0.001027 & C &  2.00$_{-0.06}^{+0.23}$ &  38.65 &3 & 7.42 &  6  \\
NGC 5746  & 0.005751 & C &  1.22$_{-0.39}^{+0.14}$ &  40.07 &3 & 7.49 &  5  \\
NGC 6240  & 0.024480 & C &  1.03$_{-0.15}^{+0.14}$ &  42.04 &3 & 9.11 &  5  \\
NGC 6251  & 0.024710 & C &  1.60$_{-0.22}^{+0.29}$ &  41.84 &3 & 8.73$^{f}$ &  6  \\
NGC 6500  & 0.010017 & C &  3.10$_{-1.70}^{+1.10}$ &  39.74 &8 & 8.28 &  6  \\
NGC 7130  & 0.016151 & C &  2.51$_{-0.22}^{+0.41}$ &  40.49 &3 & 7.54 &  11  \\
\hline
\end{tabular}
\begin{quote}
$^{a}$: distance of 3.5 Mpc \citep{pan06}. $^{*}$: Calculated using
the \cite{tre02} relation, otherwise stated: -- $^{b}$ stellar
kinematics; $^{c}$ reverberation mapping; $^{d}$ maser kinematics;
$^{e}$ $M_{\rm BH}-L_{\rm bulge}$ relation; $^{f}$ gas kinematics.
refs: -- (1) \cite{pan04}; (2) \cite{pan06}; (3) \cite{gon06}; (4)
\cite{sat05}; (5) Hypercat database at http://leda.univ-lyon1.fr/;
(6) \cite{mer03}; (7) \cite{flo06}; (8) \cite{ter03}; (9)
\cite{don06}; (10) \cite{dud05}; (11) \cite{gu06}.
\end{quote}
\end{minipage}
\end{table*}

\clearpage

\begin{thebibliography}{}

\bibitem[Bian(2005)]{bia05} Bian W. H., 2005, \cjaa, 5, 289
\bibitem[Cappi et al.(2006)]{cap06} Cappi M. et al., 2006, \aap, 446, 459
\bibitem[Carrillo et al.(1999)]{car99} Carrillo R., Masegosa J.,
Dultzin-Hacyan D., Ordo\~{n}ez R., 1999, Rev. Mex. Astron.
Astrofis., 35, 187
\bibitem[Chiaberge et al.(2006)]{chi06} Chiaberge M., Gilli R., Macchetto F. D., Sparks W. B., 2006, \apj, 651, 728
\bibitem[Dong \& De Robertis(2006)]{don06} Dong X. Y., De Robertis M. M., 2006, \aj, 131, 1236
\bibitem[Dudik et al.(2005)]{dud05} Dudik R. P., Satyapal S., Gliozzi M., Sambruna R. M., 2005, \apj, 620, 113
\bibitem[Elvis et al.(1994)]{elv94} Elvis M., Wilkes B. J., McDowell J.
C. et al., 1994, \apjs, 95, 1
\bibitem[Esin et al.(1997)]{esi97} Esin A. A., McClintock J. E., Narayan
R., 1997, \apj, 489, 865
\bibitem[Fabbiano et al.(2003)]{fab03} Fabbiano G., Elvis M., Markoff S. et al., 2003, \apj, 588, 175
\bibitem[Fabbiano et al.(2004)]{fab04} Fabbiano G., Baldi A., Pellegrini S. et al., 2004, \apj, 616, 730
\bibitem[Fabian (2004)]{fabi04} Fabian A. C., 2004, in Coevolution
of Black Holes and Galaxies, ed. L. C. Ho, 446
\bibitem[Falcke et al.(2004)]{fal04} Falcke H., K\"{o}rding E.,
Markoff S., 2004, \aap, 414, 895
\bibitem[Fanaroff \& Riley(1974)]{fan74} Fanaroff B. L., Riley J. M., 1974, \mnras, 167, 31
\bibitem[Ferrarese \& Merritt(2000)]{fer00} Ferrarese L., Merritt D., 2000, \apj, 539, L9
\bibitem[Flohic et al.(2006)]{flo06} Flohic H. M. L. G., Eracleous M., Chartas G., Shields J. C., Moran
E. C., 2006, \apj, 647, 140
\bibitem[Gliozzi et al.(2003)]{gli03} Gliozzi M., Sambruna R. M., Brandt W. N., 2003, \aap, 408, 949
\bibitem[Gliozzi et al.(2008)]{gli08} Gliozzi M., Foschini L., Sambruna R. M., Tavecchio
F., 2008, \aap, 478, 723
\bibitem[Greene \& Ho(2007)]{gre07} Greene J. E., Ho L. C., 2007, \apj, 656, 84 
\bibitem[Gonz\'{a}lez-Mart\'{i}n et al.(2006)]{gon06} Gonz\'{a}lez-Mart\'{i}n O.,
Masegosa J., M\'{a}rquez I., Guerrero M. A., Dultzin-Hacyan D.,
2006, \aap, 460, 45
\bibitem[Gu et al.(2006)]{gu06} Gu Q., Melnick J., Fernandes R. C. et al., 2006, \mnras, 366, 480
\bibitem[Haardt \& Maraschi(1991)]{haa91} Haardt F., Maraschi L., 1991, \apj, 380, L51
\bibitem[Haardt \& Maraschi(1993)]{haa93} Haardt F., Maraschi L., 1993, \apj, 413, 507
\bibitem[Heckman(1980)]{hec80} Heckman T. M., 1980, \aap, 87, 152
\bibitem[Ho et al.(1993)]{ho93} Ho L. C., Filippenko A. V., Sargent W. L. W., 1993, \apj, 417, 63
\bibitem[Ho et al.(1995)]{ho95} Ho L. C., Filippenko A. V., Sargent W. L. W., 1995, \apjs, 98, 477
\bibitem[Ho et al.(1997)]{ho97} Ho L. C., Filippenko A. V.,
Sargent W. L. W., 1997, \apjs, 112, 315
\bibitem[Ho(1999)]{ho99} Ho L. C., 1999, \apj, 516, 672
\bibitem[Ho (2008)]{ho08} Ho L. C., 2008, \araa, 46,475
\bibitem[Hopkins et al.(2007)]{hop07} Hopkins P. F., Richards G. T., Hernquist
L., 2007, \apj, 654, 731
\bibitem[Kelly et al.(2008)]{kel08} Kelly B. C., Bechtold J., Trump J. R., Vestergaard M., Siemiginowska A., \apjs, 2008, 176, 355
\bibitem[Kewley et al.(2006)]{kew06} Kewley L. J., Groves B., Kauffmann G., Heckman T., 2006, \mnras, 372, 961
\bibitem[Lewis, Eracleous \& Michael(2003)]{lew03} Lewis K. T., Eracleous M., Sambruna R. M., 2003, \apj, 593, 115
\bibitem[Lu \& Yu(1999)]{lu99} Lu Y. J., Yu Q. J., 1999, \apj, 526, L5
\bibitem[Maoz(2007)]{mao07}Maoz D., 2007, \mnras, 377, 1696 
\bibitem[Marconi et al.(2004)]{mar04}Marconi A., Risaliti G., Gilli R. et al., 2004, \mnras, 351, 169
\bibitem[Merloni et al.(2003)]{mer03} Merloni A., Heinz S., di Matteo
T., 2003, \mnras, 345, 1057
\bibitem[Narayan \& Yi(1994)]{nar94} Narayan R., Yi I., 1994, \apj, 428, L13
\bibitem[Narayan \& Yi(1995)]{nar95} Narayan R., Yi I., 1995, \apj, 452, 710
\bibitem[Narayan et al.(1998)]{nar98} Narayan R., Mahadevan R., Quataert E., 1998, in The Theory of Black
Hole Accretion Discs, ed. M. A. Abramowicz, G. Bjornsson \& J. E.
Pringle (Cambridge: Cambridge Univ. Press), 148
\bibitem[Papadakis et al.(2008)]{pap08} Papadakis I. E., Ioannou Z., Brinkmann W., Xilouris E.
M., 2008, \aap, 490, 995
\bibitem[Panessa \& Bassani(2002)]{pan02} Panessa F., Bassani L., 2002, \aap, 394, 435
\bibitem[Panessa(2004)]{pan04} Panessa F., 2004, Ph.D. Thesis,
University of Bologna, (http://venus.ifca.unican.es/$\sim$panessa/)
\bibitem[Panessa et al.(2006)]{pan06} Panessa F., Bassani L.,
Cappi M. et al., 2006, \aap, 455, 173
\bibitem[Pellegrini et al.(2003)]{pel03} Pellegrini S., Baldi A., Fabbiano G., Kim D.-W., 2003, \apj, 585, 677
\bibitem[Peterson et al.(2004)]{pet04} Peterson B. M. et al., 2004, ApJ, 613, 682
\bibitem[Raymond \& Smith(1977)]{ray77} Raymond J. C., Smith B. W., 1977, ApJS, 35, 419
\bibitem[Rybicki \& Lightman(1979)]{ryb79} Rybicki G. B., Lightman A. P., 1979, Radiation Processes in Astrophysics
(New York: Wiley)
\bibitem[Satyapal et al.(2005)]{sat05} Satyapal S., Dudik R. P., O'Halloran B., Gliozzi
M., 2005, \apj, 633, 86
\bibitem[Shakura \& Sunyaev(1973)]{sha73} Shakura N. I., Sunyaev R. A., 1973, \aap, 24, 337
\bibitem[Shemmer et al.(2006)]{she06} Shemmer O., Brandt W. N., Netzer H., Maiolino R., Kaspi
S., 2006, \apj, 646, L29
\bibitem[Shemmer et al.(2008)]{she08} Shemmer O., Brandt W. N., Netzer H., Maiolino R., Kaspi S., 2008, \apj, 682, 81
\bibitem[Terashima \& Wilson(2003)]{ter03} Terashima Y., Wilson A. S., 2003, \apj, 583, 145
\bibitem[Tremaine et al.(2002)]{tre02} Tremaine S., Gebhardt K., Bender R. et al., 2002, \apj, 574, 740
\bibitem[Vasudevan \& Fabian(2007)]{vas07} Vasudevan R. V., Fabian A. C., 2007, \mnras, 381, 1235
\bibitem[Vasudevan \& Fabian(2009)]{vas09} Vasudevan R. V., Fabian A. C., 2009, MNRAS, 392, 1124
\bibitem[Veilleux \& Osterbrock(1987)]{vei87} Veilleux S., Osterbrock D. E., 1987, \apjs, 63, 295
\bibitem[Wang et al.(2004)]{wan04} Wang J. M., Watarai K. Y., Mineshige S., 2004, \apj, 607, L107
\bibitem[Wu et al.(2007)]{wu07}Wu Q. W., Yuan F., Cao X. W.,
2007, \apj, 669, 96 
\bibitem[Yamaoka et al.(2005)]{yam05} Yamaoka K., Uzawa M., Arai M., Yamazaki T., Yoshida
A., 2005, \cjaa, 5, 273
\bibitem[Yuan et al.(2002)]{yua02} Yuan F., Markoff S., Falcke H., Biermann P. L., 2002, \aap, 391, 139
\bibitem[Yuan \& Cui(2005)]{yua05} Yuan F., Cui W., 2005, \apj, 629, 408
\bibitem[Yuan(2007)]{yua07a} Yuan F., 2007, in ASP Conf. Ser. 373, The Central Engine of Active Galactic
Nuclei, ed. L. C. Ho \& J.-M. Wang (San Francisco: ASP), 95
\bibitem[Yuan et al.(2007)]{yua07b} Yuan F., Taam R. E., Misra R., Wu X. B., Xue
Y. Q., 2007, \apj, 658, 282 

\end{thebibliography}
\end{document}